\documentstyle[12pt]{article}
\newcommand{\nc}{\newcommand}
\nc{\renc}{\renewcommand}

\nc{\etal}{\mbox{\it et al. }}
\nc{\ie}{{\it i.e.\ }}
\nc{\eg}{{\it e.g.}}

\renc{\thefootnote}{\arabic{footnote}}
\nc{\capt}[1]{{\bf Figure.} {\small\sl #1}}

\nc{\eqs}[2]{\mbox{Eqs.~(\ref{#1},\,\ref{#2})}}
\nc{\eq}[1]{\mbox{Eq.~(\ref{#1})}}

\nc{\figs}[2]{\mbox{Figs.~(\ref{#1},\,\ref{#2})}}
\nc{\fig}[1]{\mbox{Fig~.(\ref{#1})}}

\nc{\tag}[1]{\label{#1} \marginpar{{\footnotesize #1}}}
\nc{\mtag}[1]{\label{#1} \mbox{\marginpar{{\footnotesize #1}}}}
\renc{\baselinestretch}{1.2}
\newlength{\overeqskip}
\newlength{\undereqskip}
\nc{\be}[1]{\begin{equation} \mbox{$\label{#1}$}}
\nc{\bea}[1]{\begin{eqnarray} \mbox{$\label{#1}$}}
\nc{\Section}[2]{\section{#2}\label{#1}}
\nc{\Bibitem}[1]{\bibitem{#1}}
\nc{\Label}[1]{\label{#1}}

\nc{\eea}{\vspace{\undereqskip}\end{eqnarray}}
\nc{\ee}{\vspace{\undereqskip}\end{equation}}
\nc{\bdm}{\begin{displaymath}}
\nc{\edm}{\end{displaymath}}
\nc{\dpsty}{\displaystyle}
\nc{\bc}{\begin{center}}
\nc{\ec}{\end{center}}
\nc{\ba}{\begin{array}}
\nc{\ea}{\end{array}}
\nc{\bab}{\begin{abstract}}
\nc{\eab}{\end{abstract}}
\nc{\btab}{\begin{tabular}}
\nc{\etab}{\end{tabular}}
\nc{\bit}{\begin{itemize}}
\nc{\eit}{\end{itemize}}
\nc{\ben}{\begin{enumerate}}
\nc{\een}{\end{enumerate}}
\nc{\bfig}{\begin{figure}}
\nc{\efig}{\end{figure}}
\nc{\arreq}{&\!=\!&}
\nc{\arrmi}{&\!-\!&}
\nc{\arrpl}{&\!+\!&}
\nc{\arrap}{&\!\!\!\approx\!\!\!&}
\nc{\non}{\nonumber\\*}
\nc{\align}{\!\!\!\!\!\!\!\!&&}

\def\lsim{\; \raise0.3ex\hbox{$<$\kern-0.75em
      \raise-1.1ex\hbox{$\sim$}}\; }
\def\gsim{\; \raise0.3ex\hbox{$>$\kern-0.75em
      \raise-1.1ex\hbox{$\sim$}}\; }
\nc{\DOT}{\hspace{-0.08in}{\bf .}\hspace{0.1in}}
\nc{\Laada}{\hbox {$\sqcap$ \kern -1em $\sqcup$}}
\nc\loota{{\scriptstyle\sqcap\kern-0.55em\hbox{$\scriptstyle\sqcup$}}}
\nc\Loota{{\sqcap\kern-0.65em\hbox{$\sqcup$}}}
\nc\laada{\Loota}
\nc{\qed}{\hskip 3em \hbox{\BOX} \vskip 2ex}

\nc{\real}{{\rm I \! R}}
\nc{\Z}{{\sf Z \!\!\! Z}}
\nc{\complex}{{\rm C\!\!\! {\sf I}\,\,}}
\def\bigid{\leavevmode\hbox{\small1\kern-3.8pt\normalsize1}}
\def\id{\leavevmode\hbox{\small1\kern-3.3pt\normalsize1}}
\nc{\slask}{\!\!\!/}
\nc{\bis}{{\prime\prime}}
\nc{\pa}{\partial}
\nc{\na}{\nabla}
\nc{\ra}{\rangle}
\nc{\la}{\langle}
\nc{\goto}{\rightarrow}
\nc{\swap}{\leftrightarrow}

\nc{\EE}[1]{ \mbox{$\cdot10^{#1}$} }
\nc{\abs}[1]{\left|#1\right|}
\nc{\at}[2]{\left.#1\right|_{#2}}
\nc{\norm}[1]{\|#1\|}
\nc{\abscut}[2]{\Abs{#1}_{\scriptscriptstyle#2}}
\nc{\vek}[1]{{\rm\bf #1}}
\nc{\integral}[2]{\int\limits_{#1}^{#2}}
\nc{\inv}[1]{\frac{1}{#1}}
\nc{\dd}[2]{{{\partial #1}\over{\partial #2}}}
\nc{\ddd}[2]{{{{\partial}^2 #1}\over{\partial {#2}^2}}}
\nc{\dddd}[3]{{{{\partial}^2 #1}\over
	{\partial #2 \partial #3}}}
\nc{\dder}[2]{{{d #1}\over{d #2}}}
\nc{\ddder}[2]{{{d^2 #1}\over{d {#2}^2}}}
\nc{\dddder}[3]{{d^2 #1}\over
	{d #2 d #3}}
\nc{\dx}[1]{d\,^{#1}x}
\nc{\dy}[1]{d\,^{#1}y}
\nc{\dz}[1]{d\,^{#1}z}
\nc{\dl}[1]{\frac{d\,^{#1}l}{(2\pi)^{#1}}}
\nc{\dk}[1]{\frac{d\,^{#1}k}{(2\pi)^{#1}}}
\nc{\dq}[1]{\frac{d\,^{#1}q}{(2\pi)^{#1}}}

\nc{\cc}{\mbox{$c.c.$ }}
\nc{\hc}{\mbox{$h.c.$ }}
\nc{\cf}{cf.\ }
\nc{\erfc}{{\rm erfc}}
\nc{\Tr}{{\rm Tr\,}}
\nc{\tr}{{\rm tr\,}}
\nc{\pol}{{\rm pol}}
\nc{\sign}{{\rm sign}}
\nc{\bfT}{{\bf T }}
\def\GeV{{\rm\ GeV}}

\nc{\cA}{{\cal A}}
\nc{\cB}{{\cal B}}
\nc{\cD}{{\cal D}}
\nc{\cE}{{\cal E}}
\nc{\cG}{{\cal G}}
\nc{\cH}{{\cal H}}
\nc{\cL}{{\cal L}}
\nc{\cO}{{\cal O}}
\nc{\cT}{{\cal T}}
\nc{\cN}{{\cal N}}
\nc{\rvac}[1]{|{\cal O}#1\rangle}
\nc{\lvac}[1]{\langle{\cal O}#1|}
\nc{\rvacb}[1]{|{\cal O}_\beta #1\rangle}
\nc{\lvacb}[1]{\langle{\cal O}_\beta #1 |}
\nc{\bb}{\bar{\beta}}
\nc{\bt}{\tilde{\beta}}
\nc{\ctH}{\tilde{\cal H}}
\nc{\chH}{\hat{\cal H}}
\nc{\1}{\aa}
\nc{\2}{\"{a}}
\nc{\3}{\"{o}}
\nc{\4}{\AA}
\nc{\5}{\"{A}}
\nc{\6}{\"{O}}
\nc{\al}{\alpha}
\nc{\g}{\gamma}
\nc{\Del}{\Delta}
\nc{\e}{\epsilon}
\nc{\eps}{\epsilon}
\nc{\lam}{\lambda}
\nc{\om}{\omega}
\nc{\Om}{\Omega}
\nc{\ve}{\varepsilon}
\nc{\mn}{{\mu\nu}}
\nc{\k}{\kappa}
\nc{\vp}{\varphi}

\nc{\advp}[3]{{\it  Adv.\ in\ Phys.\ }{{\bf #1} {(#2)} {#3}}}
\nc{\annp}[3]{{\it  Ann.\ Phys.\ (N.Y.)\ }{{\bf #1} {(#2)} {#3}}}
\nc{\apl}[3]{{\it  Appl. Phys. Lett. }{{\bf #1} {(#2)} {#3}}}
\nc{\apj}[3]{{\it  Ap.\ J.\ }{{\bf #1} {(#2)} {#3}}}
\nc{\apjl}[3]{{\it  Ap.\ J.\ Lett.\ }{{\bf #1} {(#2)} {#3}}}
\nc{\app}[3]{{\it Astropart.\ Phys.\ }{{\bf #1} {(#2)} {#3}}}  
\nc{\cmp}[3]{{\it  Comm.\ Math.\ Phys.\ }{{ \bf #1} {(#2)} {#3}}}
\nc{\cqg}[3]{{\it  Class.\ Quant.\ Grav.\ }{{\bf #1} {(#2)} {#3}}}
\nc{\epl}[3]{{\it  Europhys.\ Lett.\ }{{\bf #1} {(#2)} {#3}}}
\nc{\ijmp}[3]{{\it Int.\ J.\ Mod.\ Phys.\ }{{\bf #1} {(#2)} {#3}}}
\nc{\ijtp}[3]{{\it Int.\ J.\ Theor.\ Phys.\ }{{\bf #1} {(#2)} {#3}}}
\nc{\jmp}[3]{{\it  J.\ Math.\ Phys.\ }{{ \bf #1} {(#2)} {#3}}}
\nc{\jpa}[3]{{\it  J.\ Phys.\ A\ }{{\bf #1} {(#2)} {#3}}}
\nc{\jpc}[3]{{\it  J.\ Phys.\ C\ }{{\bf #1} {(#2)} {#3}}}
\nc{\jap}[3]{{\it J.\ Appl.\ Phys.\ }{{\bf #1} {(#2)} {#3}}}
\nc{\jpsj}[3]{{\it J.\ Phys.\ Soc.\ Japan\ }{{\bf #1} {(#2)} {#3}}}
\nc{\lmp}[3]{{\it Lett.\ Math.\ Phys.\ }{{\bf #1} {(#2)} {#3}}}
\nc{\mpl}[3]{{\it  Mod.\ Phys.\ Lett.\ }{{\bf #1} {(#2)} {#3}}}
\nc{\ncim}[3]{{\it  Nuov.\ Cim.\ }{{\bf #1} {(#2)} {#3}}}
\nc{\np}[3]{{\it  Nucl.\ Phys.\ }{{\bf #1} {(#2)} {#3}}}
\nc{\pr}[3]{{\it Phys.\ Rev.\ }{{\bf #1} {(#2)} {#3}}}
\nc{\pra}[3]{{\it  Phys.\ Rev.\ A\ }{{\bf #1} {(#2)} {#3}}}
\nc{\prb}[3]{{\it  Phys.\ Rev.\ B\ }{{{\bf #1} {(#2)} {#3}}}}
\nc{\prc}[3]{{\it  Phys.\ Rev.\ C\ }{{\bf #1} {(#2)} {#3}}}
\nc{\prd}[3]{{\it  Phys.\ Rev.\ D\ }{{\bf #1} {(#2)} {#3}}}
\nc{\prl}[3]{{\it Phys\ Rev.\ Lett.\ }{{\bf #1} {(#2)} {#3}}}
\nc{\pl}[3]{{\it  Phys.\ Lett.\ }{{\bf #1} {(#2)} {#3}}}
\nc{\prep}[3]{{\it Phys\. Rep.\ }{{\bf #1} {(#2)} {#3}}}
\nc{\prsl}[3]{{\it Proc.\ R.\ Soc.\ London\ }{{\bf #1} {(#2)} {#3}}}
\nc{\ptp}[3]{{\it  Prog.\ Theor.\ Phys.\ }{{\bf #1} {(#2)} {#3}}}
\nc{\ptps}[3]{{\it  Prog\ Theor.\ Phys.\ suppl.\ }{{\bf #1} {(#2)} {#3}}}
\nc{\physa}[3]{{\it  Physica\ A\ }{{\bf #1} {(#2)} {#3}}}
\nc{\physb}[3]{{\it  Physica\ B\ }{{\bf #1} {(#2)} {#3}}}
\nc{\phys}[3]{{\it Physica\ }{{\bf #1} {(#2)} {#3}}}
\nc{\rmp}[3]{{\it  Rev.\ Mod.\ Phys.\ }{{\bf #1} {(#2)} {#3}}}
\nc{\rpp}[3]{{\it Rep.\ Prog.\ Phys.\ }{{\bf #1} {(#2)} {#3}}}
\nc{\sjnp}[3]{{\it Sov.\ J.\ Nucl.\ Phys.\ }{{\bf #1} {(#2)} {#3}}}
\nc{\spjetp}[3]{{\it Sov.\ Phys.\ JETP\ }{{\bf #1} {(#2)} {#3}}}
\nc{\yf}[3]{{\it Yad.\ Fiz.\ }{{\bf #1} {(#2)} {#3}}}
\nc{\zetp}[3]{{\it Zh.\ Eksp.\ Teor.\ Fiz.\  }{{\bf #1}  {(#2)} {#3}}}
\nc{\zp}[3]{{\it Z.\ Phys.\ }{{\bf #1} {(#2)} {#3}}}
\nc{\ibid}[3]{{\sl ibid.\ }{{\bf #1} {#2} {#3}}}
\nc{\rf}[1]{(\ref{#1})}
\nc{\nn}{\nonumber \\*}
\nc{\bfB}{\bf{B}}
\nc{\bfv}{\bf{v}}
\nc{\bfx}{\bf{x}}
\nc{\bfy}{\bf{y}}
\nc{\vx}{\vec{x}}
\nc{\vy}{\vec{y}}
\nc{\oB}{\overline{B}}
\nc{\oI}{\overline{I}}
\nc{\oR}{\overline{R}}
\nc{\rar}{\rightarrow}
\nc{\ti}{\times}
\nc{\slsh}{\hskip-5pt/}
\nc{\sm}{Standard~Model~}
\nc{\w}{\omega}
\nc{\MP}{M_{\rm Pl}}
\nc{\tp}{t_{\rm Pl}}
\nc{\ave}{\bar{E}}

\renc{\min}{p_{\rm min}}
\renc{\max}{p_{\rm max}}
\nc{\pmin}{p_{\rm min}}
\nc{\pmax}{p_{\rm max}}
\nc{\fo}{f_0}
\nc{\foi}{f_{0,i}\,}
\nc{\fop}{f_0^P}
\nc{\fou}{f_0^U}
\def\sepand{\rule{14cm}{0pt}\and}
\nc{\eff}{{\rm eff}}
\nc{\MT}{M_{\rm T}}
\nc{\ML}{M_{\rm L}}
\nc{\kk}{\vek{k}}
\nc{\pp}{{\rm p}}
\nc{\cb}{critical bubble~}
\nc{\cbs}{critical bubbles~}
\nc{\scb}{subcritical bubble~}
\nc{\scbs}{subcritical bubbles~}
\nc{\gta}{\hspace{3pt}^>\hspace{-7pt}{\scriptstyle\sim}\hspace{3pt}}
\begin{document}

{\title{{\hfill {{\small  TURKU-FL-P35-00
        }}\vskip 1truecm}
{\bf Q-ball collisions in the MSSM: gauge-mediated 
supersymmetry breaking}}

 
\author{
{\sc Tuomas Multam\" aki$^{1\dagger}$}\\
{\sl and}\\
{\sc Iiro Vilja$^{2}$ }\\ 
{\sl Department of Physics,
University of Turku} \\
{\sl FIN-20014 Turku, Finland} \\
\sepand
}
\date{May 17th, 2000}
\maketitle}

\vspace{2cm}
\begin{abstract}
\noindent 
Collisions of non-topological solitons, Q-balls, are considered
in the Minimal Supersymmetric Standard Model where supersymmetry 
has been broken at a low energy scale via a gauge mediated mechanism.
Q-ball collisions are studied numerically on a two dimensional lattice 
for a range of Q-ball charges. Total cross-sections, as well as
fusion and geometrical cross-sections are calculated. The total and 
geometrical cross-sections appear to converge with increasing charge. The 
fusion cross-section has been estimated to be larger than 60\% of 
the geometrical cross-section for large balls.

\end{abstract}
\vfill
\footnoterule
{\small$^1$tuomas@maxwell.tfy.utu.fi,  $^2$vilja@newton.tfy.utu.fi\\}
{\small $\dagger$ Supported by the Finnish Graduate School in Nuclear and 
Particle Physics}
\thispagestyle{empty}
\newpage
\setcounter{page}{1}
\section{Introduction}
Various field theories can support stable non-topological solitons
\cite{leepang}, Q-balls \cite{coleman}. 
A Q-ball is a coherent scalar
condensate that carries a conserved charge, typically a $U(1)$ charge.
Due to charge conservation, the Q-ball configuration is the ground state
in the sector of fixed charge. 
Q-balls may have physical importance because the
supersymmetric extensions of the Standard Model have
scalar potentials that are suitable for Q-balls to exist in the theory. 
In particular, lepton or baryon number carrying Q-balls
are present in the Minimal Supersymmetric Standard Model (MSSM) due 
to the existence of flat directions in the scalar sector
of the theory \cite{kusennko2,enqvist1}.

If supersymmetry (SUSY) is broken at low energy scales
by a gauge mediated mechanism, the scalar potential 
is completely flat for large enough field values.
Therefore the energy per unit charge decreases like $B^{-1/4}$
($B$ is the baryon number) and for large enough $B$ one can have 
completely stable B-balls since there are no light enough baryon number
carrying particles the Q-ball can decay into \cite{kusenko3}.
If SUSY breaking is due to a hidden supergravity sector
in the theory, the potential is not flat. Radiative
corrections allow for $Q$-balls to exists but
they are unstable and decay typically to quarks and nucleons
\cite{enqvist1, enqvistdm}. The decay progresses by
evaporation from the surface of the Q-ball \cite{cohen, multamak}.

Q-balls can be cosmologically significant in various ways. 
Stable (or long living) Q-balls are natural candidates
for dark matter \cite{kusenko3}. Their decay offers 
a way to understand the baryon to dark matter ratio
\cite{enqvistdm} and the baryon asymmetry of the universe
\cite{enqvist1}. Q-balls can also protect the baryons
from electroweak sphalerons \cite{enqvist1} and may
be an important factor in considering the stability of 
neutron stars \cite{wreck}.

For Q-balls to be cosmologically significant one needs
to have a mechanism that creates them in early stages
of the evolution of the universe. Q-balls can be created
in the early universe from an Affleck-Dine (AD) condensate 
\cite{enqvist1,kusenko3,enqvistdm}. This process has been
studied recently by numerical simulations \cite{kawasaki1, kawasaki2}
where both the gauge- and gravity-mediated SUSY breaking scenarios
were considered. In both cases Q-balls with various charges
were seen to form from the condensate.

Collisions of Q-balls have been considered previously
in various potentials \cite{kawasaki2}-\cite{multamak2}.
In \cite{kawasaki2} collisions were simulated on a one dimensional lattice 
in the gravity-mediated case and it was found that Q-balls typically 
merge, exchange charge or pass through each other \cite{kawasaki2}.
To our knowledge, collisions
have not been studied in the gauge-mediated case previously
and the gravity-mediated case has been studied in more than one
dimension only in \cite{multamak2}.
In a recent paper Q-ball collisions were studied
numerically in one, two and three dimensions in a polynomial potential
\cite{battye}. The main qualitative features of the collision processes 
were similar in all the three cases. 

Since the charge of Q-ball can change in a collision process,
they may play an important role in the evolution of 
the Q-ball charge distribution after their formation. 
On the other hand, the Q-ball charge distribution affects
the cosmological role that Q-balls may have in the evolution of the universe.
The effect of collisions can therefore be an important factor 
in evaluating the cosmological role of Q-balls.

In this paper we have studied Q-ball collisions 
in the gauge-mediated scenario on a two dimensional lattice. 
The gravity-mediated scenario has been analyzed in a previous paper
\cite{multamak2}.

\section{Q-ball solutions}
Consider a field theory with a U(1) symmetric scalar potential,
$U(\phi)$, with a global minimum at $\phi=0$. The complex scalar field
$\phi$ carries a unit quantum number with respect to the $U(1)$-symmetry.
The charge and energy of a field configuration $\phi$ 
in D dimensions are \cite{leepang}
\be{charge}
Q={1\over i}\int (\phi^*\partial_t\phi-\phi\partial_t\phi^*)d^Dx
\ee
and
\be{energy}
E=\int [|\dot{\phi}|^2+|\nabla\phi|^2+U(\phi^*\phi)]d^Dx.
\ee

The single Q-ball solution is the minimum energy configuration 
in the sector of fixed charge. The Q-ball will be stable
against radiative decays into $\phi$-scalars if condition
\be{stabcond}
E<mQ,
\ee
where $m$ is the mass of the $\phi$-scalar, holds. It is then
energetically favourable to store charge in a Q-ball rather
than in form of free scalars.

Finding the minimum energy is straightforward using
Lagrange multipliers. The Q-ball can be shown to be of the form
\cite{leepang}
\be{ansatz}
\phi(x,t)=e^{i\omega t}\phi(r),
\ee
where $\phi(x)$ is now time independent and real, $\omega$ is 
the Q-ball frequency, $|\omega|\in[0,m]$ and $\phi$ is spherically
symmetric.

The charge of a Q-ball with spherical symmetry in D-dimensions is
given by
\be{sphercharge}
Q=2\omega\int \phi(r)^2 d^Dr
\ee
and the equation of motion at a fixed $\omega$ is 
\be{eom}
{d^2\phi\over dr^2}+{D-1\over r}{d\phi\over dr}=\phi{\partial
U(\phi^2)\over\partial\phi^2}-\omega^2\phi.
\ee 
To find the Q-ball solution we must solve (\ref{eom}) with
the boundary conditions $\phi'(0)=0,\ \phi(\infty)=0$.

In the present paper we consider a potential of the form
 \cite{kusenko3}
\be{potential}
U(\phi) =m^4(1+\log({\phi^2\over m^2}))+{\lambda^2\over
M^2} \phi^{6},
\ee
with parameter values $m=10^4$ GeV, $\lambda=0.5$ and 
$M=2.4\times 10^{18}$ GeV. This corresponds to a potential
along a flat direction that has been lifted by soft 
supersymmetry-breaking terms. Supersymmetry is broken here
by a gauge-mediated mechanism as opposed to the
gravity-mediated case analyzed previously \cite{multamak2}.

\begin{figure}[ht]
\leavevmode
\centering
\vspace*{55mm}
\includegraphics{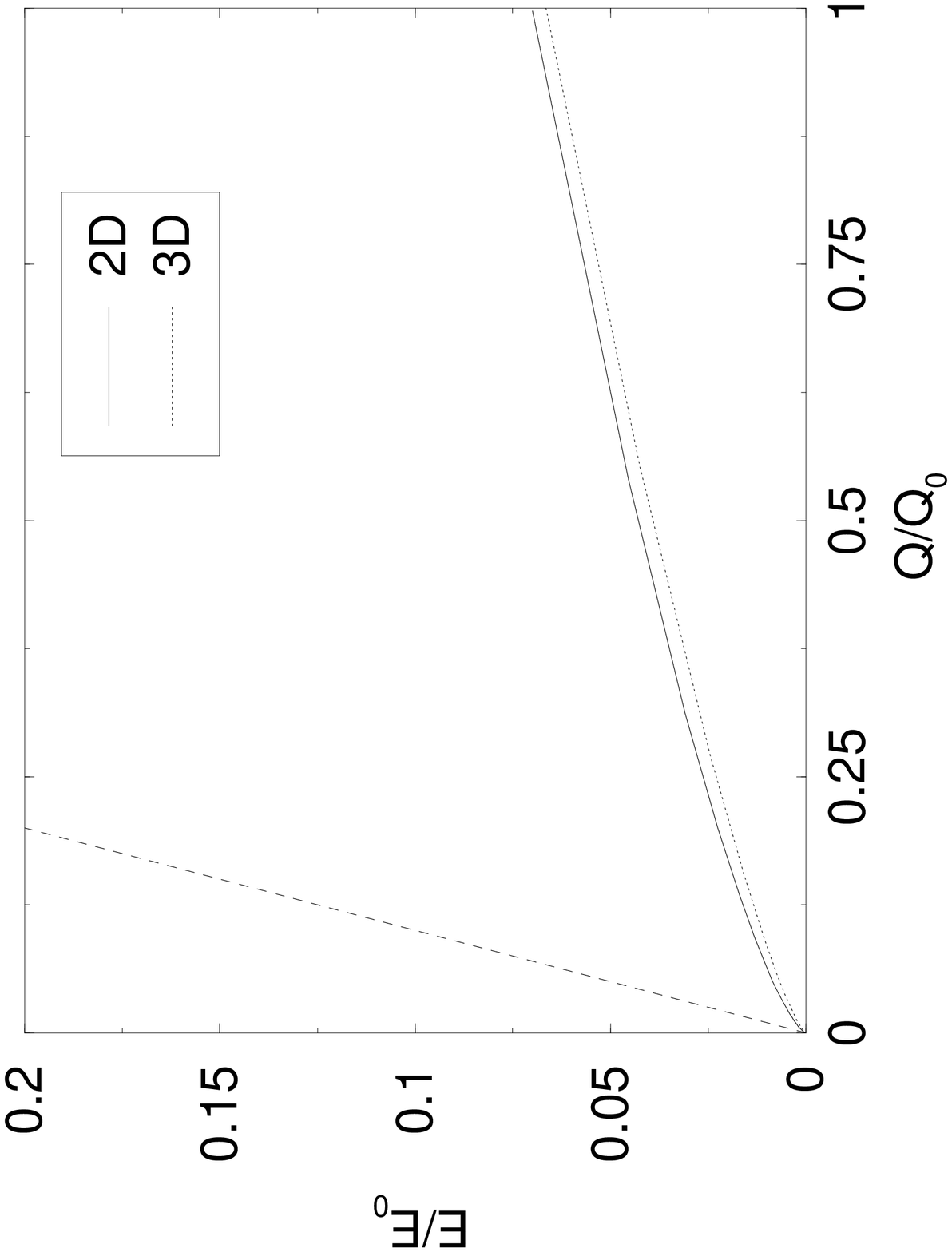}
\includegraphics{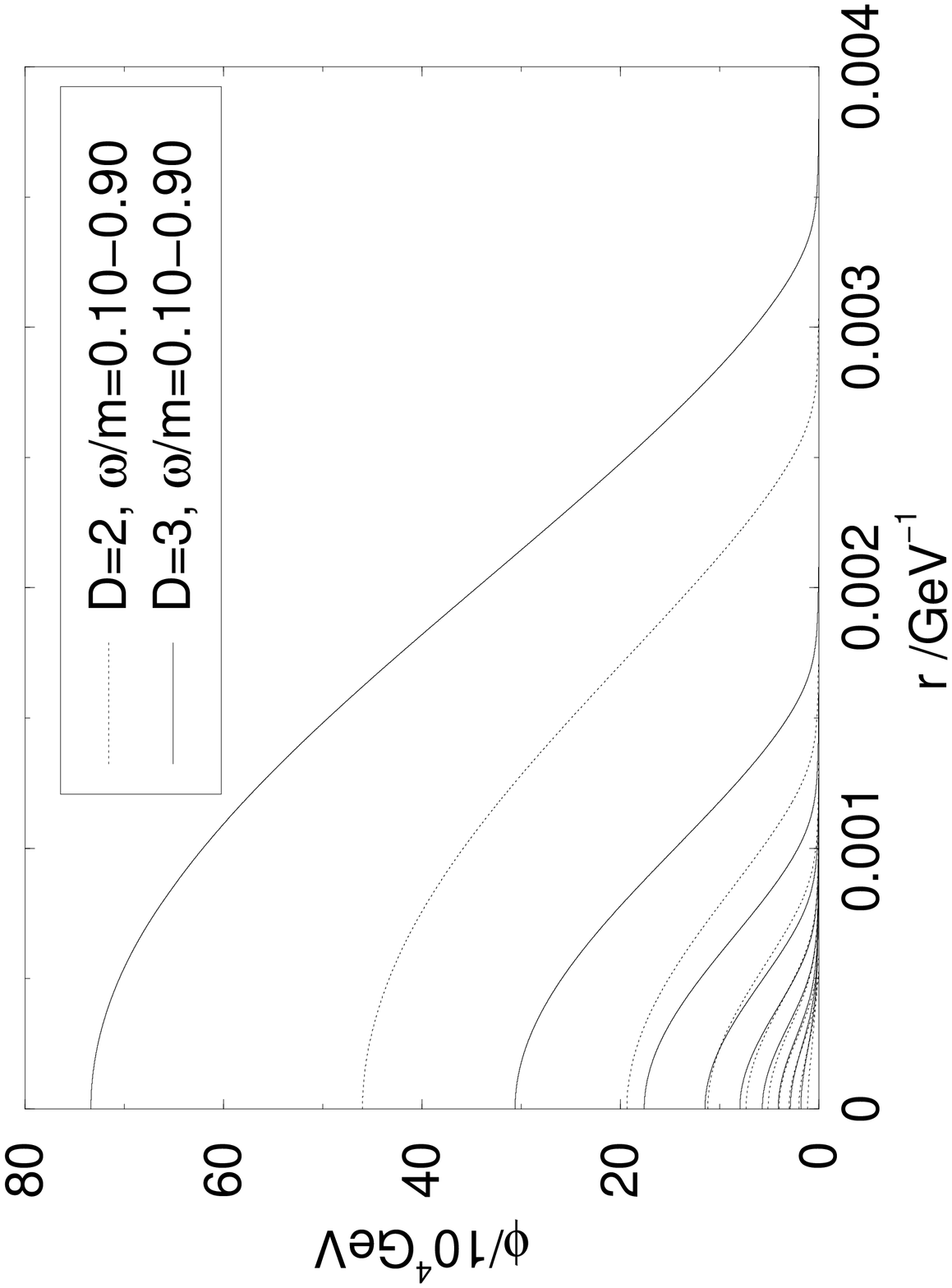}

\hspace{1cm}(a)\hspace{7cm}(b)
\caption{Q-ball energy as a function of charge and Q-ball profiles
in two and three dimensions.}
\label{evsq}
\end{figure}

We have calculated the charge and energy of Q-balls for different values
of $\w$. Energy vs.\ charge curves are shown in Figure \ref{evsq}(a).
The axis scales are chosen differently for two and three dimensions; 
for two dimensions, $Q_0=240(m/\GeV)^2$, $E_0=mQ_0\GeV$
and for three dimensions, $Q_0=4.0(m/\GeV)^2$, $E_0=mQ_0\GeV$.
The dashed line is the stability line, $E=mQ$, that 
indicates that the Q-balls considered here are stable with
respect to scalar decays.

As from Fig. \ref{evsq}(a) can be seen the two and three dimensional 
energy vs. charge curves follow each other very closely.
Also the Q-ball profiles in two and three dimensions have very 
similar shapes, Fig. \ref{evsq}(b). The similarities in the
energy vs. charge curves and Q-ball profiles gives an indication
that collisions studied in two and three dimensions are likely
to give similar results. In \cite{battye} it was also found that
the two and three dimensional cases possess similar qualitative features.
In contrast, the one dimensional case
is fundamentally different from the higher dimensional cases.
As from (\ref{eom}) can be seen, there is no dissipation
term in one spatial dimension. Processes can then have 
properties that are not seen in higher dimensions. In the
one dimensional simulations we have studied this also seemed to be the
case. 

Even though the general features of collision processes are similar
in different potentials, the exact form of the potential
is still important. From the basis of our simulations it appears
that the details of collisions are dependent on the choice of potential.

One should also point out that in three dimensions new, ring-like 
intermediate states were seen in collisions at high velocities
\cite{battye}. The corresponding process in two dimensions is
a right angle scattering -process that we did not observe in
our simulations. However, at high velocities, $\cO(0.1)$, there can be
new types of processes that we have not seen in our low velocity
simulations.

\section{Collisions}
We have studied collisions of Q-balls with equal charges
in the potential (\ref{potential}).
The range of $\w$ for which numerical simulations have been done
is $\w/m=0.15,\ 0.20,\ 0.30,\ 0.60$. In terms of charges these
values of $\w$ in the two dimensional case correspond to $5.6, 2.1,
0.47, 0.032$ (in units of $({m\over\GeV})^2$). 
The initial velocities of the balls are allowed to have two values,
$v=10^{-3}$ or $v=10^{-2}$. 

Collisions are studied over different relative phases,
$0\leq\Delta\phi\leq\pi$. Here we have defined the relative phase,
$\Delta\phi$, to be the difference in individual Q-ball phases at the 
point where the distance between them would be at a minimum assuming 
there were no interaction between them.  If the balls 
are of equal size, the point at which the relative phase difference
is defined is irrelevant. In general, however, one needs to define
the phase difference so that it is independent of the initial positions
of the balls.

The position of a Q-ball is defined by the location of its maximum amplitude.
We have varied the impact parameter to study 
the scattering cross-sections. As from Fig. \ref{evsq} can be seen, the 
Q-balls studied here have thick walls so that one needs a definition
for their size. In the gravity mediated case \cite{multamak2} we defined 
the size
of a ball by a Gaussian fit. Here, however, we have found that the
Gaussian is a much poorer fit than in the gravity-mediated case.
Instead we use a kink solution, $\phi=A+B\tanh(Cr+D)$,
and find that it approximates the numerical profiles well.
The radius of the ball is defined as $R={1\over C}(\tanh^{-1}(-{A\over B})-D)$.
This definition has the advantage that as the profile of the ball
approaches the thin-wall solution, radius becomes defined in a natural
way. It is worth noting that even though the profiles
visually appear to be similar to the Q-ball profiles in the gravity-mediated
scenario, they are fundamentally different from them and approach
a purely thin-walled profile in the large charge limit.

A two dimensional, typically a $\sim 300\times 300$, lattice 
with continuous boundary conditions was used in all calculations.

\subsection{Numerical Results}

As in \cite{multamak2}, the collision processes 
that we have observed can be roughly divided into three categories; 
fusion, charge
exchange and elastic scattering. Fusion is defined as a process where
most of the initial charge is in a single Q-ball after the collision 
and the rest of the charge is lost either as radiation or as small Q-balls.
By charge exchange we mean a process where Q-balls exchange
some of their charge while the total amount of charge carried by the
two balls is essentially conserved. An elastic scattering is defined
to be a scattering process where less than $1\%$ of the total charge 
is exchanged.

The type of a collision process is mainly dependent on 
the relative phase difference, $\Delta\phi$, between the colliding
Q-balls. When $\Delta\phi$ is small
the Q-balls fuse and form a larger ball. Excess charge
is lost in form of radiation and small lumps of charge. As $\Delta\phi$
increases the Q-balls no longer fuse and start to scatter while exchanging 
a significant amount of their
charge. The amount of charge that is exchanged in a collision
decreases with increasing phase difference until the Q-balls
scatter elastically. The change from
a fusion process to an elastic scattering is typically rapid so that a 
significant amount of charge is exchanged only at a very narrow range
of $\Delta\phi$. This is a different behavior from the gravity-mediated
case where charge exchange was generally a much more dominant process
\cite{multamak2}.
The qualitative effects of
changing the relative phase difference are the same for the
whole range of $\w$:s and initial velocities that we have studied. 

The effect of the impact parameter is much less pronounced than
the phase difference on the type of collision. Typically if the balls
fuse at zero impact parameter, they continue to fuse with an increasing
impact parameter until at some point the balls start to scatter
elastically or while exchanging little of their charge. 
The interaction probability, averaged over the different phases,
also has a clear cut-off with respect to the impact parameter.

From the simulations we can now calculate the total, fusion and
geometric cross-sections averaged over the relative phases.
The charge exchange cross-section is typically much smaller than the
other cross-sections and is not quoted here.
The total cross-section includes all the different types
of processes. The quoted cross-sections are the three dimensional 
cross-sections with the interaction radius taken from the two dimensional
simulations.

\begin{figure}[ht]
\leavevmode
\centering
\vspace*{110mm}
\includegraphics{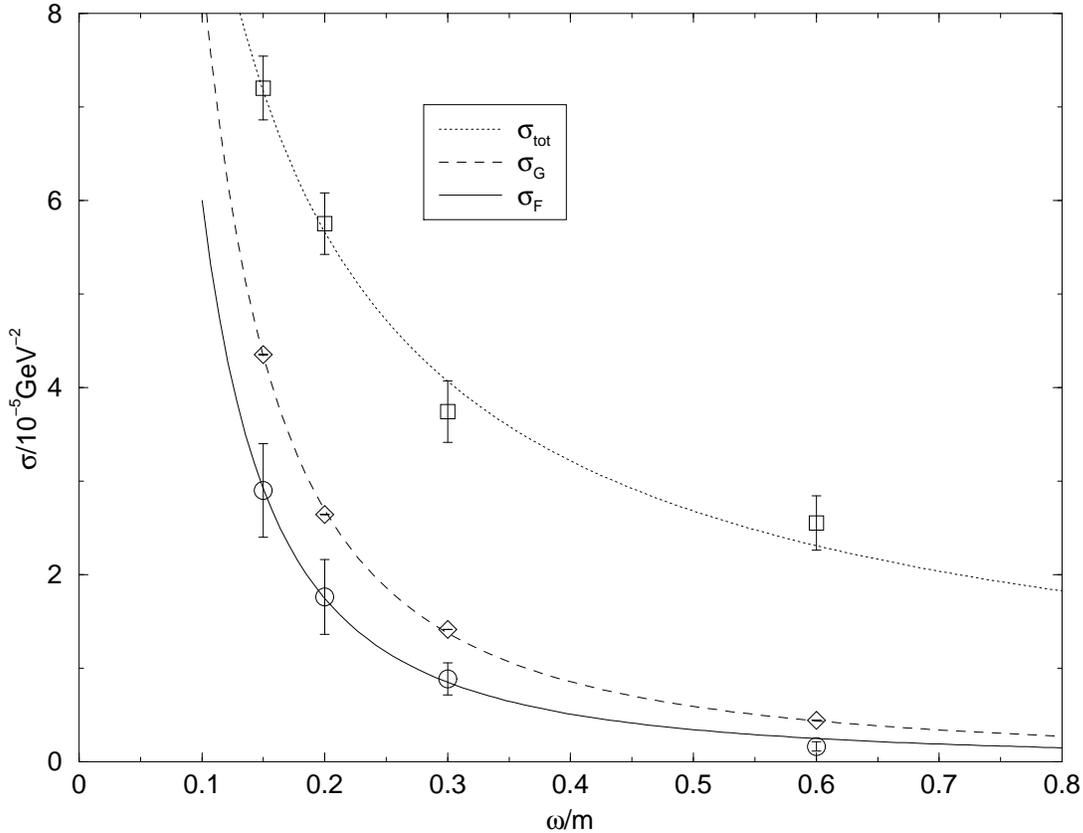}
\caption{The geometric, total and fusion
cross-sections for different values of $\w$, $v=10^{-3}$.}
\label{sigmas1}
\end{figure} 

\begin{figure}[ht]
\leavevmode
\centering
\vspace*{110mm}
\includegraphics{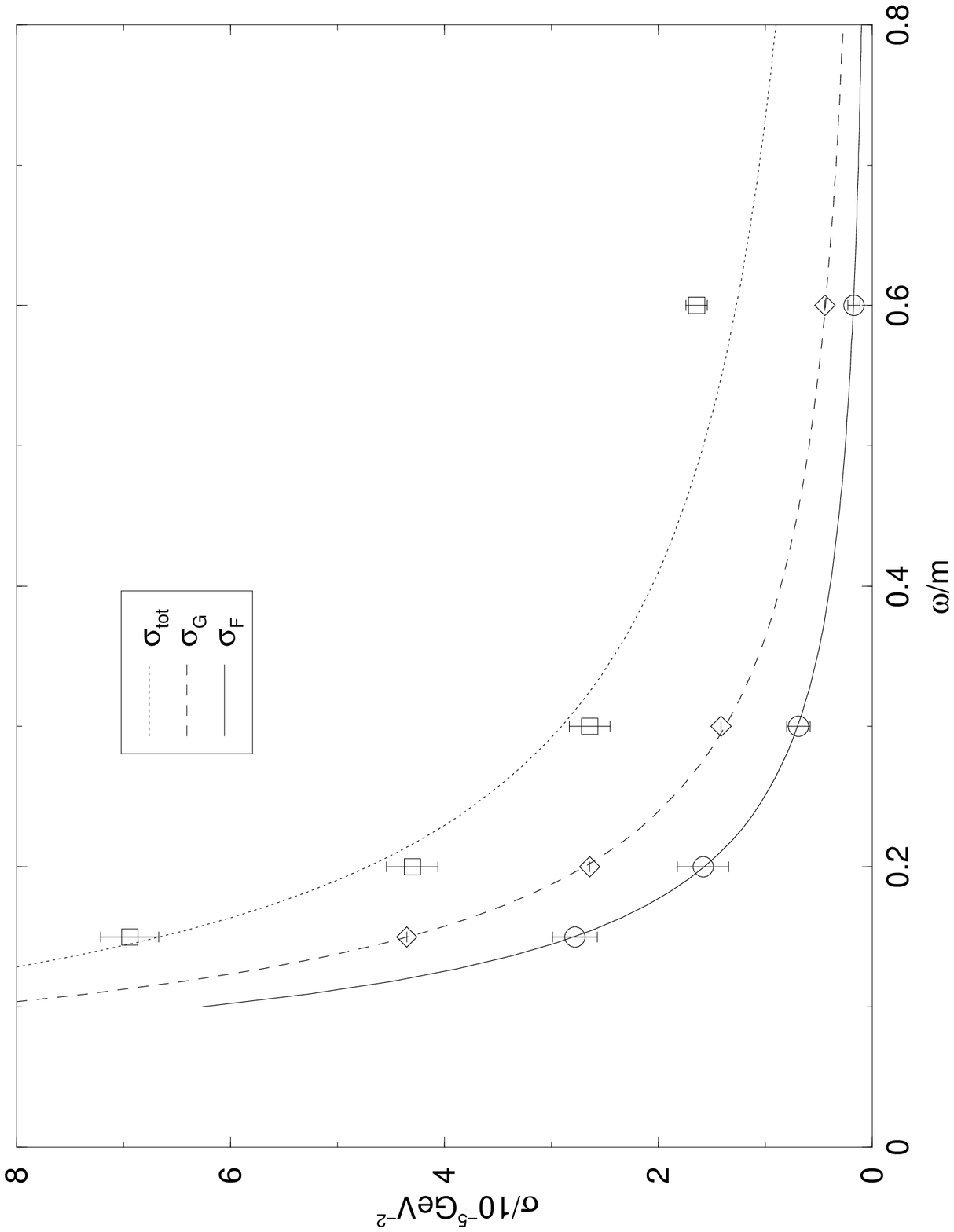}
\caption{The geometric, total and fusion
cross-sections for different values of $\w$, $v=10^{-2}$.}
\label{sigmas2}
\end{figure} 

The geometric, total and fusion cross-sections for 
$v=10^{-3}$ are plotted in Fig. \ref{sigmas1}
and in Fig. \ref{sigmas2} for $v=10^{-2}$.
In each case we have
fitted a curve $\sigma=A(\w/m)^B$ through the points.
From the figures it is clear that the cross-sections increase with decreasing
$\w$. This is naturally due to the increasing size
of the Q-balls with decreasing $\w$.
More importantly, one can also see that the ratio of the total
cross-section to the geometrical cross-section decreases with
decreasing $\w$ \ie increasing charge. This reflects the fact 
that as the profile of the Q-ball approaches the thin-wall type,
the geometrical cross-section is an increasingly better approximation
to the total cross-section. This is intuitively clear; as the
balls become more and more thin-walled, the effects of the boundary 
diminish and the geometrical cross-section dominates the total
cross-section.
The ratio of the fusion cross-section 
to the geometrical cross-section is increasing, but only very
slightly, with increasing charge. Extrapolating to large, thin-walled
balls one can therefore conclude that the total cross-section
of large Q-balls is well approximated by their geometrical
cross-sections. The fusion cross-section of such balls can then be 
bounded by a fraction of the geometrical cross-section,
$\sigma_F/\sigma_G\gsim 0.6$. The geometrical cross-sections of large 
Q-balls can be estimated analytically, $R\approx
{1\over\sqrt{2}}m^{-1}Q^{1/4}$ \cite{kusenko4} so that $\sigma_G\approx
2\pi m^{-2}Q^{1/2}$.

The effect of the increased velocity can also be seen from the
figures. As the initial velocity is increased tenfold to $10^{-2}$,
the total and fusion cross-sections both decrease. This is expected
since faster balls have less time to interact with each other 
and are more likely to pass without interacting.

Compared to the gravity-mediated case \cite{multamak2}, the
results presented here show both some similarities and some differences.
In the gravity-mediated case the radius of the balls, and hence the
geometrical cross-section, was approximately constant whereas
here the radius of the ball varies greatly which obviously has an effect
on the different cross-sections. Also the probabilities of different
types of processes are different in the two scenarios: in the gravity-mediated 
case a charge exchange-process is much more likely to occur than a
fusion process, here the charge exchange cross-section is much smaller.
The fusion processes also can be distinguished in the two cases,
in the gravity-mediated case charge was typically lost as small
Q-balls whereas here most of the charge is lost as radiation and
lumps of charge. This demonstrates clearly that the exact form of the
potential is significant even though the general qualitative
features are alike.
One can also spot similarities between the gravity- and gauge-mediated
scenarios; in both cases the fusion cross-sections increase with 
increasing balls and an increase in velocity decreases the total
cross-sections. 

\section{Conclusions}

In this paper we have studied Q-ball collisions in the MSSM
with SUSY broken by a gauge mediated mechanism.
It was found that Q-balls may fuse, exchange charge or
scatter elastically in a collision depending on the
relative phase difference between them. The probability
of each process is dependent not only on the relative
phase difference but also on the size of the balls.
Larger balls are more probable to fuse in a collision
whereas smaller balls are more likely to scatter
either elastically or while exchanging some of their charge.
Collisions can therefore alter the charge distribution of Q-balls
quite significantly, provided that collisions are frequent 
enough.

Our simulations give an indication that the total cross-section,
$\sigma_{\mathrm{tot}}$,
approaches the geometrical cross-section, $\sigma_G$, as the Q-ball size grows
so that in the thin-wall limit the total and geometrical cross-sections 
are equal. The fusion cross-section also grows with the ball size
and on the basis of our results we can give an estimate for the fusion
cross-section of large balls, $\sigma_F \gsim 0.6\ \sigma_G$.

If collisions are to play a significant role in cosmology, 
the interaction rate must be large enough at some time in the evolution
of the universe. If the average interaction rate is smaller than the
Hubble rate after the Q-balls are formed, the distribution will 'freeze
out' and collisions will not alter the charge distribution.
If, on the other hand, the collision rate is initially larger than 
the Hubble rate, collisions can affect the charge distribution until
the distribution freezes out due to the expansion of the universe.
Furthermore, if fusion processes dominate, the number density of
Q-balls may decrease rapidly which can also freeze the
Q-ball charge distribution.

On the basis of our results, the size and the relative phase
differences are important in determining the evolution of the
Q-ball charge distribution in the gauge mediated scenario.
If the Q-balls are initially in the same phase, they typically
fuse in a collision. The average size of a Q-ball then increases
while the number density decreases. As our results show, 
the fusion probability increase with increasing Q-ball size
so that larger balls are more likely to increase their size
even further by collisions. The Q-ball interaction rate is dependent 
on several factors; clearly the 
initial Q-ball charge distribution, the interaction cross-section, 
number density and the velocity distribution affects the average
interaction rate. 

Q-balls may play a significant role in the past and present stages
of the universe. To be able to address the question of their
significance more decisively one needs to consider not only
the initial charge distribution but also its evolution. Collisions
may be an important factor and to study their effect in more
detail poses an interesting question that motivates further
study.

\vspace{1cm}
\noindent{\bf Acknowledgements.}
We thank K. Enqvist for discussions and the Center for Scientific
Computing for computation resources. This work has been supported 
by the Finnish Graduate School in Nuclear and Particle Physics
and by the Academy of Finland under the project no. 40677.


\end{document}